\documentclass[twocolumn,aps,prb,showpacs,superscriptaddress,unsortedaddress]{revtex4}
\usepackage{graphicx}
\usepackage{epsf}
\newcommand{\etal}{{\it et al.}}

\begin{document}

\title{Modeling the Fermi arc  in underdoped cuprates}

\author{M. R. Norman}
\affiliation{Materials Science Division, Argonne National Laboratory, Argonne, IL 60439}
\author{A. Kanigel}
\affiliation{Department of Physics, University of Illinois at Chicago, Chicago, IL 60607}
\author{M. Randeria}
\affiliation{Department of Physics, The Ohio State University, Columbus, OH  43210}
\author{U. Chatterjee}
\affiliation{Department of Physics, University of Illinois at Chicago, Chicago, IL 60607}
\author{J. C. Campuzano}
\affiliation{Department of Physics, University of Illinois at Chicago, Chicago, IL 60607}
\affiliation{Materials Science Division, Argonne National Laboratory, Argonne, IL 60439}

\begin{abstract}
Angle resolved photoemission data in the pseudogap phase of underdoped cuprates have 
revealed the presence of a truncated Fermi surface consisting of Fermi arcs.  We compare
a number of proposed models for the arcs, and find that the one that best models the data  
is a d-wave energy gap with a lifetime broadening whose
temperature dependence is suggestive of fluctuating pairs.
\end{abstract}

\date{\today}
\maketitle

\section{Introduction}

It is well established that cuprates possess a superconducting phase with an order
parameter that has d-wave symmetry,\cite{Tsuei} and for hole-doped materials,
this phase exists over a range of doping above 5\%.
It is also well established that at very low dopings, the material is an
antiferromagetic Mott insulator.  Connecting these two states is an unusual phase
known as the pseudogap, the nature of which is still being debated.\cite{NPK}
It is felt by many that the proper identification of this phase will have a major
impact on the ultimate `mechanism' for pairing in cuprate superconductors.

Angle resolved photoemission spectroscopy (ARPES) reveals the presence of a truncated
Fermi surface in the pseudogap phase.\cite{Marshall,Ding,Loeser}
In a study of the pseudogap versus temperature,\cite{Nature98} this truncated Fermi surface
was denoted as a `Fermi arc'.  The arc was shown to be intermediate between the d-wave node
of the superconductor and the complete Fermi surface of the normal state.
Moreover, the arc appears to form by a closing of the energy gap of the
superconducting state as the temperature is raised above T$_c$.  Off the arc,
in the `pseudogapped' (antinodal) region of the Brillouin zone, the spectral gap appears instead
to fill in with temperature.  This filling in effect is also seen in c-axis conductivity data \cite{Homes}
and is consistent with the thermal evolution of the specific heat.\cite{Loram}
This `dual' nature of the energy gap is suggestive of a `two gap'
scenario where a `superconducting' gap resides on the arc and a `pseudogap' resides off
the arc.  Such a two gap picture was proposed by Deutscher,\cite{Deutscher}
and recent Raman,\cite{Sacuto} ARPES,\cite{Tanaka,Kondo} and STM \cite{Hudson}
data  have been offered in its support.  On the other hand, even for underdoped samples,
the gap function below T$_c$ seems to be more or less d-wave like.\cite{Mesot99}
This conundrum of having a single gap below T$_c$ transforming into a dual gap above T$_c$
was stressed sometime ago.\cite{Norman99}

Recently, a very detailed temperature and doping study of the energy gap above T$_c$
was done by Kanigel \etal\cite{NatPhys}  They found that the length of the arc scales
as T/T*, where T*, the temperature at which the spectral
gap `fills up' in the antinodal region of the zone, strongly increases with 
underdoping.\cite{JC99}  As a consequence, the angular anisotropy of the pseudogap
looks more and more like a d-wave gap as the temperature is lowered relative to T*.
This finding is supported by thermal conductivity data, which indicates that the d-wave dispersion
of the superconducting state at low temperatures survives when the doping is reduced
into the pseudogap state.\cite{Doiron}  It is also consistent with recent ARPES and STM
data on the stripe ordered phase of La$_{7/8}$Ba$_{1/8}$CuO$_4$, which indicates
a d-wave like gap anisotropy at low temperatures (but above T$_c$).\cite{Valla}  More recently, the study
of Kanigel \etal\ has been extended to below T$_c$,\cite{AmitPRL} where it was found that
the arc collapses to a node within the resistive width of the transition, with a simple d-wave
like gap below T$_c$.  These recent studies bring into question the `two gap' picture.

Primarily motivated by the ARPES data, a wide range of models have been proposed to
explain the Fermi arc.  Basically, these models can be grouped into two categories.  In the
first, the pseudogap is associated with a q=0 instability.  Most of the models in this category
have the pseudogap as a precursor to the superconducting gap, and involve pair formation
with the absence of long range phase order.\cite{Randeria,Kivelson95,Varenna}  These models have
been extended to describe the arc by explicitly invoking vortex-like
excitations \cite{Franz,Zlatko,PWA,Berg} as revealed by measurements of the Nernst
effect.\cite{Ong}  The node of the d-wave dispersion is broadened into an arc by a combination
of lifetime broadening as well as Doppler shifts of the single particle states
due to the vortices.  There are, though, q=0
theories which do not involve superconductivity.  One example is the model of Varma 
and Zhu,\cite{Varma}
which involves circulating currents within a CuO$_2$ plaquette (and thus has the same periodicity
as the unit cell).  In this case, the gap function  has a `d$^2$' anisotropy.  Another example is
the `nodal nematic' phase of Kim \etal,\cite{Kim} where the node is displaced by a nematic
order parameter rather than a vortex Doppler shift.  The final example we mention
is the model of Wen and
Lee \cite{Wen} where the node is displaced in energy due to staggered flux phase
correlations.  In fact, a rich variety of behavior has been predicted within the general context
of resonating valence bond (RVB) theories.\cite{LeeRev, Gros,Sensarma,Para}

The second category involves a non-zero q vector.  This category ranges from models based on
a precursor spin density wave,\cite{Tremblay, Kuch}
charge density wave,\cite{LZW} stripes,\cite{Granath}
flux phases,\cite{LeeRev} or orbital currents.\cite{DDW} In the case of
fluctuating order,\cite{LeeRev,Civelli,Kaul,Guidry} the non-zero q vector is not as
obvious in the excitation spectrum.  Those scenarios involving a $(\pi,\pi)$
wavevector possess small hole pockets centered at $(\pi/2,\pi/2)$
where the intensity is reduced on one side of the pocket due to
the amplitude factors which mix the states differing by q.
Related models are those where the Luttinger
surface (surface of zeros of the single particle Greens function) differs from the Fermi
surface.\cite{Yang,Stanescu07,Phillips,Stanescu06}  In this case, the Fermi surface is
truncated where it crosses the Luttinger surface.  In a more general 2k$_F$ context, the
flat parts of the Fermi surface which reside in the antinodal region of the zone
can be eliminated by nesting,\cite{Tsvelik,McElroy} leaving a residual arc.

In this paper, some of these scenarios will be addressed in the context of the ARPES
data.  In Section II, several
non-zero q scenarios, where for simplicity long range order is assumed, will be analyzed.
These scenarios typically lead to (a) Fermi arcs whose length is T independent,
(b) deviations of the arcs from the underlying Fermi surface,
(c) energy gaps which are not centered symmetrically about the Fermi energy, and
(d) shadow bands.
We argue that there is no evidence for these
effects in ARPES and tunneling data, at least in the mildly underdoped region.
In Section III, we turn to the q=0 solutions.  We find that the
scenario most consistent with the data is one where the node remains along the zone
diagonal and at the Fermi energy.  The temperature evolution of the arc above T$_c$ is consistent
with lifetime broadening of the node, though the data also indicate a distortion of the 
d-wave gap anisotropy 
with temperature.
%This lifetime, though, is not the `single particle' lifetime, but
%rather the `pair' lifetime associated with the gap contribution to the self-energy.
%This is most apparent from the evolution of the arc through T$_c$.
In Section IV, we offer some  conclusions,
and suggest future ARPES experiments that could further differentiate between the various models
for the Fermi arc.
%We also comment on the implications of the arc in regards to transport data.

\section{Non zero q scenarios}

\subsection{Commensurate density wave}

These scenarios assume a q vector of $(\pi,\pi)$ with an energy
gap that is either isotropic, or has d-wave symmetry.\cite{DDW}
The secular matrix is of 2 by 2 form, and the Greens function associated with the wavevector k
in the presence of simple elastic broadening, $\Gamma$, can be written as:
\begin{eqnarray}
G_k & = & \left(\frac{E_+ - \epsilon_{k+q}}{E_+-E_-}\right)\frac{1}{\omega-E_++i\Gamma} \\ \nonumber
& & -\left(\frac{E_- - \epsilon_{k+q}}{E_+-E_-}\right)\frac{1}{\omega-E_-+i\Gamma}
\end{eqnarray}
where
\begin{equation}
E_{\pm} = \frac{\epsilon_k+\epsilon_{k+q}}{2}
\pm \sqrt{\left(\frac{\epsilon_k-\epsilon_{k+q}}{2}\right)^2+\Delta_k^2}
\end{equation}
We have looked at several cases, with various dispersions, $\epsilon_k$, including some with
bilayer splitting, and several different forms for $\Delta_k$.  For brevity, we present results using
for $\epsilon_k$ the tight binding dispersion of Norman \etal\cite{Norman95} and a d-density
wave gap \cite{DDW} $\Delta_k=\frac{\Delta_0}{2}(\cos(k_x)-\cos(k_y))$.

\begin{figure}
\centerline{\includegraphics[width=3.4in]{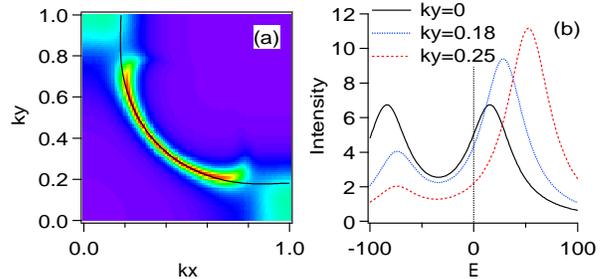}}
\caption{(Color online) (a) Spectral intensity at zero energy versus $k_x, k_y$, and
(b) versus energy for $k_x=1$ for several $k_y$, for the d-density wave
model.\cite{DDW}  The black curve in (a) is the normal state Fermi surface.
Zone dimensions for all figures are in $\pi/a$
units, and energies are in meV.  For all figures (unless otherwise noted), $\Delta_0$=50 meV
and $\Gamma$=25 meV.}
\label{fig1}
\end{figure}

In Fig.~1a, we present the intensity plot of the spectral function
(imaginary part of $G_k$) in the 2D zone 
for $\omega=0$.  At the simplest level,
one indeed finds an arc.  But there are several details worth
pointing out.  First, the ends of the arc turn away from the underlying Fermi surface of the normal
state.  This is due to the fact that the zero energy contour traces out a pocket centered at $(\pi/2,\pi/2)$,
the back side of which is suppressed by the coherence factors (the prefactor of each term in Eq.~1).
Second, there is a strong suppression of the intensity at the `hot spots' - the points where the normal
state Fermi surface ($\epsilon_k=0$) crosses its $(\pi,\pi)$ displaced image.  This can be related to the `Luttinger
surface' effect mentioned in  the Introduction.  To see this, we note that the Greens function in this
model can be rewritten as:
\begin{equation}
G^{-1}_k = \omega - \epsilon_k + i\Gamma -\frac{\Delta_k^2}{\omega-\epsilon_{k+q}+i\Gamma}
\end{equation}
The `gap' self-energy (the last term of this equation) diverges when 
$\omega=\epsilon_{k+q}$
in the absence of broadening ($\Gamma=0$).
Thus the $(\pi,\pi)$ translated image of the normal state Fermi surface ($\epsilon_{k+q}=0$) is the
Luttinger surface, and therefore the zero energy intensity is suppressed when the normal
state Fermi surface crosses
this surface.  Finally, there is weaker intensity centered around the $(\pi,0)$ points which will be
suppressed as $\Delta_0$ increases.  To investigate this further,
in Fig.~1b, we show the spectral function for several k points along the $(\pi,0)-(\pi,\pi)$ direction.
One clearly sees that the spectral function has a minimum value that sits at negative energy.
At $k=(\pi,0)$, it is obvious from Eq.~2 that this minimum value occurs at $\omega=\epsilon_k$,
which is -34 meV for this dispersion. This asymmetry in energy is obviously enhanced for 
dispersions where
$\epsilon_{\pi,0}$ is deeper in energy.

\begin{figure}
\centerline{\includegraphics[width=3.4in]{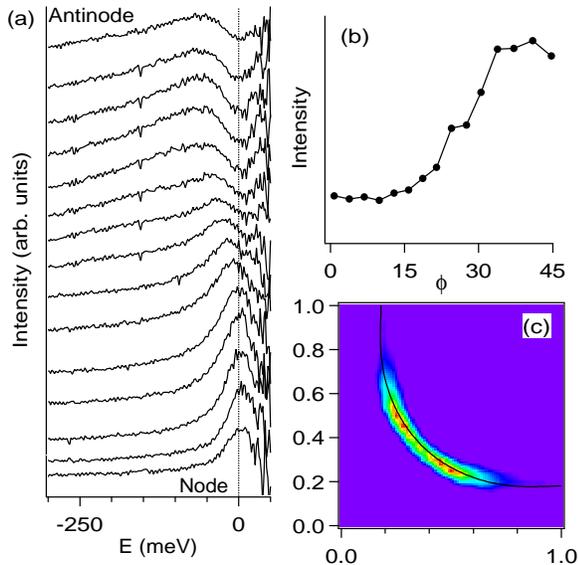}}
\caption{(Color online) (a) Experimental energy distribution curves (EDCs) for optimal doped 
Bi$_2$Sr$_2$CaCu$_2$O$_8$ (Bi2212) around the underlying
Fermi surface in the pseudogap phase (T=140K) divided by a resolution broadened Fermi
function.  The bottom curve is at the node and the top curve at the antinode.  The data
set is the same as in Fig.~1b of Ref.~\onlinecite{NatPhys}.
(b) Zero energy intensity from (a) as a function of the Fermi surface angle, $\phi$,
where $\phi=0^{\circ}$ corresponds to the antinode and $\phi=45^{\circ}$ to the node.
(c) Zero energy intensity versus $k_x, k_y$ (the data were reflected relative to $k_x=k_y$). 
For (b), the intensities were obtained by subtracting a background EDC
(obtained from an unoccupied k),
then normalizing this subtracted intensity by its energy integrated weight.
This was designed to minimize the effect of the photoemission matrix elements.
This was not done in (c) in order to demonstrate that the raw data show no indication for
any deviation of the arc from the underlying Fermi surface (black curve).}
\label{fig2}
\end{figure}

In relation to the experimental data, we note the following issues with this model, which are generic
to models based on a finite q order parameter.  First, there is no natural way to generate an arc
whose length is proportional to the temperature.  Second, there is no evidence from ARPES
for a `turn in' of the ends of the arc away from the underlying normal state 
Fermi surface (Fig.~2c).  Third, ARPES is consistent
with spectral functions which either have a maximum (arc) or minimum (antinodal region) at
zero energy along the
underlying Fermi surface.  We demonstrate this in Fig.~2a, where data in the pseudogap phase
along the underlying Fermi surface is plotted.  These data are the same as used to construct
Fig.~1b of Ref.~\onlinecite{NatPhys}, but instead of `symmetrizing' the raw data as was done
there (which implicitly assumes a maximum or minimum at zero energy), we divide the data
by a resolution broadened Fermi function.  The clear maxima at zero energy along the arc, and the
minima at zero energy away from the arc, are quite evident.  This is consistent with tunneling data as
well \cite{Fischer} where the minimum in the tunneling conductance is at zero bias, even in
the pseudogap phase.

\subsection{Differing Luttinger surface}

These scenarios \cite{Yang,Stanescu07,Phillips,Stanescu06} are related to the ones 
just discussed.  For discussion purposes, we look at the recently proposed model of
Yang, Rice, and Zhang.\cite{Yang}  In this case, the Greens function is
\begin{equation}
G^{-1}_k = \omega - \epsilon_k + i\Gamma -\frac{\Delta_k^2}{\omega+\epsilon_k^{NN}+i\Gamma}
\end{equation}
where $\epsilon_k^{NN}$ is just the near neighbor term of the tight binding dispersion
($\Delta_k$ has the same form as the d density-wave case).
Note that if $\epsilon_k$ only had a near neighbor term, then at half filling, this model would
be equivalent to the d density-wave model.  The similarity of these two models
can be seen in Fig.~3, where we show the
zero energy intensity plot in the 2D zone, as well as the intensity versus $\omega$ for k along
$(\pi,0)-(\pi,\pi)$.  Again, note the pronounced suppression of the intensity at the `hot spots' in Fig.~3a,
which is not evident in the data (a plot of the experimental zero energy intensity around the
Fermi energy is shown in Fig.~2b), as well as the pronounced asymmetry of the energy gap
relative to the Fermi energy in Fig.~3b.  And, as with the d density-wave model, there is no obvious
mechanism to obtain an arc proportional to T.

\begin{figure}
\centerline{\includegraphics[width=3.4in]{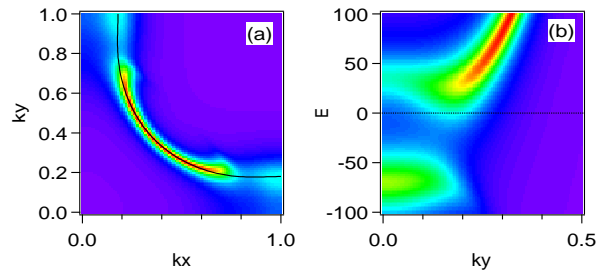}}
\caption{(Color online) (a) Spectral intensity at zero energy versus $k_x, k_y$, and
(b) versus energy and $k_y$ for $k_x=1$, for the model of
Yang \etal\cite{Yang}}
\label{fig3}
\end{figure}

\subsection{Nesting density wave}

These scenarios assume a q vector which nests the antinodal points of the 2D Fermi surface.
Two approximations were analyzed.  In the first, a single q vector along $q_y$, $q=(0,-q)$,
was used in the first octant (bounded by $(0,0)-(\pi,0)-(\pi,\pi)-(0,0)$)
of the square lattice zone (a 2 by 2 secular equation),
the result of which was then reflected to the other octant.  The orientation of q was designed so
as to connect the antinode at $(\pi,q/2)$ with the one at $(\pi,-q/2)$.
The equation for the Greens function
is the same as in Eqs.~1 and 2, except that $\Delta_k$ in this case was taken to be isotropic.

In the second approximation, a 3 by 3 secular equation is separately solved for q vectors oriented 
respectively along $q_x$, $q=(q,0)$ and $q=(-q,0)$, and along $q_y$, $q=(0,q)$ and $q=(0,-q)$,
in the first quadrant of the zone, and then the two results are averaged (representing averaging
over two domains).  The unaveraged $G_k$
is given by
\begin{equation}
G_k = \sum_{i=1}^3 \frac{(E_i - \epsilon_{k+q})(E_i - \epsilon_{k-q})}{(E_i-E_{i+1})(E_i-E_{i+2})}
\frac{1}{\omega-E_i+i\Gamma}
\end{equation}
where by $i+1$ and $i+2$ we mean modulo 3.  The $E_i$ are given by solving the appropriate
cubic equation and can be written as
\begin{equation}
E_i = -2\sqrt{d}\cos((\theta+2 \pi i)/3)-a/3
\end{equation}
where
\begin{eqnarray}
a & = & -(\epsilon_k +\epsilon_{k+q} +\epsilon_{k-q}) \nonumber \\
b & = & \epsilon_k \epsilon_{k+q} + \epsilon_k \epsilon_{k-q} + \epsilon_{k+q} \epsilon_{k-q} 
-2\Delta_k^2 \nonumber \\
c & = &  -\epsilon_k \epsilon_{k+q} \epsilon_{k-q} + \Delta_k^2 (\epsilon_{k+q} + \epsilon_{k-q})
\nonumber \\
d & = & (a^2-3b)/9 \nonumber \\
r & = & (2a^3-9ab+27c)/54 \nonumber \\
\theta & = & \cos^{-1} (r/d^{3/2})
\end{eqnarray}

In Fig.~4, we show the zero energy intensity in the 2D zone for the two approximations.  Again,
a clear arc is seen, with extra structure that can be attributed to the reduced intensity (due again
to the coherence factors) of the `shadow' bands.  This is particularly true in 4b where more shadow
bands occur.  A similar situation
would occur if one had `checkerboard' order (this would be obtained
by solving a 5 by 5 secular matrix associated with a `double q' structure).

\begin{figure}
\centerline{\includegraphics[width=3.4in]{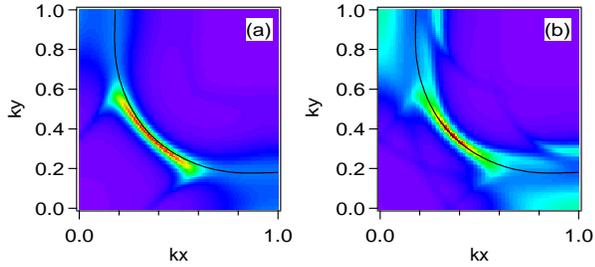}}
\caption{(Color online) (a) Spectral intensity at zero energy versus $k_x, k_y$ for (a) the 2 by 2 secular equation
and (b) the 3 by 3 secular equation approximations, for the antinodal nesting model with $q=(0.36,0)$.}
\label{fig4}
\end{figure}

A significant difference from the previous cases is the origin of the arc.
In the previous cases, the arc is due to the Fermi energy cutting across the lower of the two energy
bands.  In essence, the energy gap is centered above the Fermi energy for k vectors from the node to
the `hot spots', and it is centered below the Fermi energy for k vectors from the `hot spots' to 
the antinode.  But in this `antinodal' nesting case, it is the reverse situation.  In the 2 by 2 approximation,
the arc is formed from the Fermi energy cutting across 
the upper of the two bands.  This is particularly evident near the arc tip, as shown in Fig.~5.
In essence, the energy gap is centered below the Fermi energy for k along the arc.  
For momenta near the arc tip, one would find a minimum in the spectral function at a negative
energy (as in Fig.~1b).  This effect is not evident, though, in the ARPES data.
And again, as the arc tip
is associated with where the underlying normal state Fermi surface intersects the density wave
zone boundary (in the first octant, this would correspond to $k_y=q/2$), there is no natural
mechanism for an arc proportional to temperature.  As discussed by McElroy,\cite{McElroy} this
would require a `two gap' scenario, where the density wave gap would wipe out the antinodal parts
of the Fermi surface, and then a second gap would wipe out the remaining arc with reducing temperature.
Despite the attractiveness of such scenarios in regards to some experimental data,\cite{Deutscher,
Sacuto,Tanaka,Kondo,Hudson} a  definitive signature of this density wave gap would be to observe the
shadow bands evident in Fig.~4 and the asymmetry of the gap relative to the Fermi energy
evident in Fig.~5.  So far, we have found no evidence for either of these effects.\cite{Shadow}

\begin{figure}
\centerline{\includegraphics[width=3.4in]{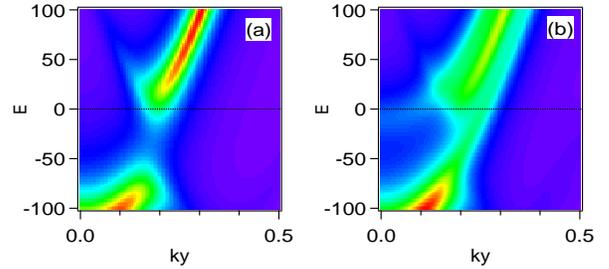}}
\caption{(Color online) (a) Spectral intensity versus energy and $k_y$ near the arc tip ($k_x=0.6$)
for (a) the 2 by 2 secular equation
and (b) the 3 by 3 secular equation approximations, for the antinodal nesting model with $q=(0.36,0)$.}
\label{fig5}
\end{figure}

\section{Zero q scenarios}

\subsection{Energy displaced node}

The RVB model of Wen and Lee \cite{Wen} is based on incorporating both the effect of a d-wave gap
in the particle-particle channel and a staggered flux phase gap in the particle-hole channel.  An
ansatz for the Greens function in this model that makes it of the same form as the earlier cases we
studied is
\begin{equation}
G^{-1}_k = \omega - \epsilon_k + i\Gamma -\frac{\Delta_k^2}{\omega+\epsilon_{k}+\mu_{sh}+i\Gamma}
\end{equation}
The effect of $\mu_{sh}$ is to move the d-wave node off the Fermi energy, and the sign of $\mu_{sh}$ is chosen
to be negative so that the node is above the Fermi energy.
The result is an arc at zero
energy that is tied to the underlying Fermi surface (Fig.~6a).    This is a positive feature of this model.
If $|\mu_{sh}|$ were proportional to T for some (unknown) reason, this could also account for the temperature
evolution of the arc.  The major problem with this model, though, is that as one goes to positive
energy, the arc would shrink in size, and as one goes to negative energy, the arc
would expand in size.  This effect is not evident in the ARPES data in the pseudogap
phase, though, which seems to be more or less consistent with arcs which are
independent of energy up to energy scales of order $\Delta_0$.\cite{Utpal06}  And
in  this model, the energy gap is always centered at an energy above the Fermi energy for all
momentum cuts in the zone, an effect not visible in Fig.~2a.

\begin{figure}
\centerline{\includegraphics[width=3.4in]{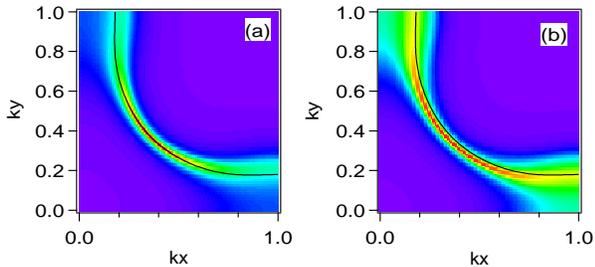}}
\caption{(Color online) (a) Spectral intensity at (a) zero energy and (b) -25 meV versus $k_x, k_y$
for the model of Wen and Lee with $\mu_{sh}$=-50meV.\cite{Wen}}
\label{fig6}
\end{figure}

\subsection{d-wave pairs plus lifetime broadening}

The simplest model in this class is equivalent to the one just described with $\mu_{sh}=0$
\begin{equation}
G^{-1}_k = \omega - \epsilon_k + i\Gamma -\frac{\Delta_k^2}{\omega+\epsilon_{k}+i\Gamma},
\end{equation}
The spectral function for
finite $\Gamma$ traces out an `arc', as shown in Fig.~7a.  And
the energy gap is centered at the Fermi energy, as shown in Fig.~7b.  That is, the
gap is tied to the Fermi energy and the Fermi surface, consistent with experiment.
In Fig.~8a, we plot the evolution of the spectral function on the Fermi surface ($\epsilon_k=0$)
for this model, and
in Fig.~8b the angular anisotropy of the spectral gap (half the peak to peak separation).

\begin{figure}
\centerline{\includegraphics[width=3.4in]{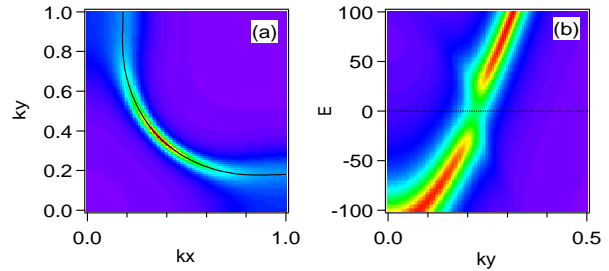}}
\caption{(Color online) (a) Spectral intensity at zero energy versus $k_x, k_y$, and
(b) versus energy and $k_y$ for $k_x=0.6$, for the d-wave pair model.}
\label{fig7}
\end{figure}

\begin{figure}
\centerline{\includegraphics[width=3.4in]{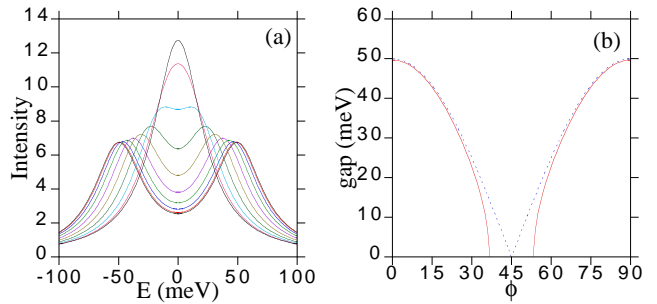}}
\caption{(Color online) (a) Spectral intensity around the Fermi surface for the d-wave pair model.
The top curve is at the node,
the bottom curve at the antinode.
(b) Spectral gap (half the peak to peak separation) versus the Fermi surface angle.
The dashed curve corresponds to $\Gamma$=0.}
\label{fig8}
\end{figure}

Gapped and ungapped spectra on the Fermi surface (Fig.~8a) are obviously controlled
by the sign of the second derivative
of the spectral function with respect to $\omega$ at $\omega=0$.  The condition that this
second derivative is zero is $\Gamma = \sqrt{3} \Delta_k$.
Assuming a simple d-wave gap of the form
$\Delta_k = \Delta_0 \cos(2\phi)$ where $\phi$ is the Fermi surface angle measured relative
to the antinode, one then obtains for the position of the arc tip
$\phi_0 = 0.5 \cos^{-1} (\Gamma/\sqrt{3}\Delta_0)$.  T* would then be the condition
that $\Gamma(T) = \sqrt{3} \Delta_0(T)$.

In Fig.~9a, we show the variation of the arc length with $\Gamma$.
This variation is consistent with experiment,
as shown in Fig.~9a, if one assumes that $\Gamma \propto T$ and $\Delta_0$ is a constant in $T$
(similar plots are shown in Refs.~\onlinecite{Guidry,Storey}).  This linear variation of the arc length
with $\Gamma$ is a natural consequence of the linear variation of $\Delta_k$ with $\phi$ around the node.
The upturn of the arc length at larger $\Gamma$ is due to the quadratic dependence
of the energy gap with $\phi$ about the antinode.

\begin{figure}
\centerline{\includegraphics[width=3.4in]{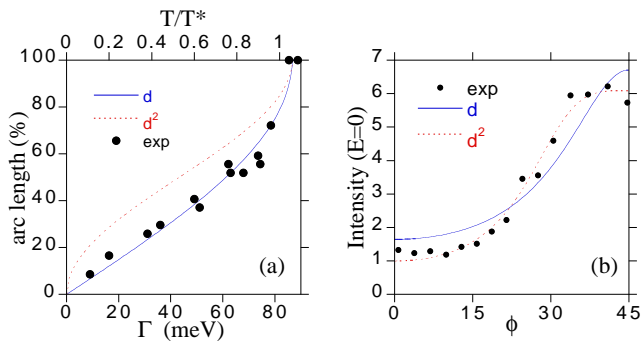}}
\caption{(Color online)
(a) Arc length versus $\Gamma$ for a d-wave gap and a $d^2$ gap
(data from Ref.~\onlinecite{NatPhys}).
(b) Experimental spectral intensity at zero energy versus Fermi surface angle (as in
Fig.~2b) versus fits that assume either a d-wave gap ($\cos(2\phi)$) or a $d^2$ gap ($\cos^2(2\phi)$).
}
\label{fig9}
\end{figure}

The theory of Varma and Zhu \cite{Varma} is similar except that $\Delta_k$ is taken to
be the square of the d-wave gap.  As they point out, this fits the angular anisotropy of
the parameter $\Delta_k$ in the pseudogap phase \cite{NatPhys} better than the simple d-wave model,
as can be seen in Fig.~9b.
On the other hand, the arc length variation with $T$ is more consistent
with the simple d-wave form, as shown in Fig.~9a, though we remark that Varma and Zhu
were able to obtain a much better fit to the arc length by allowing a self-energy with
a more sophisticated frequency and temperature dependence.\cite{Varma}

\section{Conclusions}

In regards to the `non zero q' scenarios, there are several ways that experiment could address
this question.  Definitive evidence would be finding a departure of the arc from the underlying
normal state Fermi surface, evidence for an energy gap which is asymmetric in energy
relative to the chemical potential, or the existence of shadow bands (i.e., bands displaced from
the main band by the wavevector q).  Other evidence would be the existence of intensity
suppression at `hot spots' (where the Fermi surface would cross the Luttinger surface), as has
been observed by ARPES in electron doped cuprates.\cite{NCCO}

In regards to the `zero q' scenarios,
the simplest theory consistent with the data appears to be a d-wave gap
with an inverse lifetime that is proportional to $T$.  There are, though, some limitations of
this model.  The data of Ref.~\onlinecite{NatPhys} were actually fit with a form more general
than that of Eq.~9 \cite{PRB98}
\begin{equation}
G^{-1}_k = \omega - \epsilon_k + i\Gamma_1 -\frac{\Delta_k^2}{\omega+\epsilon_{k}+i\Gamma_0}
\end{equation}
This `two lifetime' model has the advantage of being able to describe a broad spectral function
($\Gamma_1$) but with a sharp leading edge gap ($\Gamma_0$) as indicated
by ARPES data in the pseudogap phase.\cite{Ding,Nature98,PRB98}
It has since been extended to include a more general frequency dependence for the 
self-energy.\cite{PRB01}  The presence of two lifetimes may seem
unusual, as this does not occur, for instance, in the standard Eliashberg treatment of strong coupling
superconductors.\cite{Eliashberg}  The motivation in Ref.~\onlinecite{PRB98} was that
$\Gamma_1$ denotes the interaction in the particle-hole channel, whereas $\Gamma_0$
denotes that in the particle-particle channel.  If the dispersion of the single particle
states is ignored, a calculation of fermions interacting with pair fluctuations leads to 
$\Gamma_0$ proportional to $T-T_c$.\cite{PRB98}  In contrast, in
two dimensions, a dependence proportional to $\sqrt{T-T_c}$ is obtained.\cite{Maki}
Consideration of vortex excitations \cite{Franz} leads instead
to linear $T$ behavior at high $T$.
In general, one would expect linear $T$ behavior at high temperatures,
since this occurs for any model of fermions interacting with bosons,\cite{Elihu}
but with a collapse to zero at $T_c$ since the inverse pair lifetime should vanish in the 
ordered state.
This general $T$ dependence not only naturally describes the linear $T$ variation of
the arc length,\cite{NatPhys} it can also account for the `filling up' of the gap in the antinodal region.\cite{PRB98}
It also explains why the arc length collapses to zero within the resistive width of the transition.\cite{AmitPRL}
Regardless, our experience has been that two lifetimes are necessary to properly model the 
data.\cite{PRB98,PRB01,NatPhys}
A complete description of modeling based on Eq.~10
in regards to the ARPES data is beyond the scope of the present
paper and will be left for a future study.

The fits presented in Ref.~\onlinecite{NatPhys} also indicated that the gap anisotropy
changed with temperature, and that this effect could not be described by lifetime broadening
of the zero temperature gap, though it should be remarked that
the actual value of $\Delta_k$ is difficult to extract once the broadening
significantly exceeds $\Delta_k$.
Whether such changes in anisotropy (and in particular, a region around the node where
$\Delta_k$ is identically zero as indicated by these fits) can be described by 
pair breaking within a strong-coupling Eliashberg context remains to be seen.
Certainly, careful measurements of the gap anisotropy at different
temperatures and dopings would help to better differentiate models based on a d-wave gap from
more novel ones, such as that of Varma and Zhu,\cite{Varma} that have a fundamentally different gap
anisotropy.

In conclusion, we believe that a model of a d-wave gap with a temperature dependent lifetime
consistent with fluctuating pairs gives the simplest description of ARPES data in the pseudogap
phase.  Future experimental work should be aimed at further differentiating between various proposed
models for the Fermi arc, as well as using this information to address other data, such as transport.
 
\acknowledgments

This work was supported by the U.S. DOE, Office of Science, under Contract 
No.~DE-AC02-06CH11357 and by NSF DMR-0606255.


\begin{thebibliography}{99}
\bibitem{Shadow}
In Bi2212, the only observed shadow bands so far have been of structural origin, due to its
orthorhombic unit cell and its superlattice modulation with $q_{SL}=\pi(0.21,0.21)$.  The former
has a wavevector of $q_{ortho}=(\pi,\pi)$ which has complicated looking for effects
of commensurate antiferromagnetism, flux phases, and d-density waves which have the same
characteristic wavevector.  For these structural features, though, there is no evidence for an
associated energy gap.
\end{thebibliography}
\end{document}